\documentclass[a4paper]{scrartcl}

\usepackage[english]{babel}
\usepackage[utf8x]{inputenc}
\usepackage[T1]{fontenc}

\usepackage[a4paper,top=3cm,bottom=2cm,left=3cm,right=3cm,marginparwidth=1.75cm]{geometry}

\usepackage{amsmath}
\usepackage{hyperref}

\usepackage{xcolor}
\hypersetup{
    colorlinks,
    linkcolor={red!50!black},
    citecolor={blue!50!black},
    urlcolor={blue!80!black}
}

\usepackage{xpatch}

\newlength{\myspace}
\makeatletter
\xpatchcmd{\@maketitle}{\vskip.5em}{\vskip\myspace}{}{}
\makeatother

\title{Is QBism The Future of Quantum Physics?\footnote{This paper was initially published in \textit{The Quantum Times}: \href{http://www.thequantumtimes.org/2017/07/is-qbism-the-future-of-quantum-physics/}{thequantumtimes.org/2017/07/is-qbism-the-future-of-quantum-physics/}} \\ 
\large }

\subtitle{Review of Hans Christian von Baeyer's\\ \textit{QBism: The Future of Quantum Physics} [2016]}

\author{Kelvin J. McQueen\footnote{Department of Philosophy, Chapman University, Orange, CA, US.}}

\begin{document}

\maketitle

\tableofcontents

\section{Overview}

The purpose of this book is to explain Quantum Bayesianism (`QBism') to ``people without easy access to mathematical formulas and equations'' (4-5). Qbism is an interpretation of quantum mechanics that ``doesn't meddle with the technical aspects of the theory [but instead] reinterprets the fundamental terms of the theory and gives them new meaning'' (3). The most important motivation for QBism, enthusiastically stated on the book's cover, is that QBism provides ``a way past quantum theory's paradoxes and puzzles'' such that much of the weirdness associated with quantum theory ``dissolves under the lens of QBism''. 

Could a non-technical book that almost fits in your pocket really succeed in resolving the notorious paradoxes of quantum theory? I believe the answer is: not in this case. There are three primary reasons. Firstly, the argument that QBism solves quantum paradoxes is not convincing. Secondly, it proves difficult to pin down exactly what QBism says. Thirdly, as a scientific theory QBism seems explanatorily inert, which is maybe why the topic of scientific explanation is not broached. Sections 2-4 below discuss each of these problems in turn. However, there is still a wealth of insights that make this book a worthwhile read, and I shall begin with those.

The book is split into four chapters. The first presents eight sections that take the reader on a lively journey through quantum theory. This chapter will prove valuable to a variety of readers. For a general audience, it is one of the best non-technical introductions to quantum physics I have read in some time. For specialists, it illustrates clever methods for explaining quantum theory to beginners.

So much for the `Q' in QBism. Chapter two presents two sections explaining Bayesian probability theory (the `B'). The first (section 9) sets the stage by challenging common intuitions about the nature of probability e.g. using the paradox of the cube factory. It is then explained how nature can yield surprising demands on correct probability assignments. This is carefully illustrated with Bose-Einstein statistics and the exclusion principle. Finally, the probability theory known as \textit{frequentism} is sharply criticized. Section 10 espouses Bayesian probability theory. Here the central rule for rationally updating one's probability assignments in light of new evidence - Bayes' law - is explained and illustrated with care. 

My only qualm so far is that the dichotomy between frequentism and Bayesianism is slightly misleading. For much of the book suggests that adopting QBism is a ``decision to switch from the frequentist to the Bayesian interpretation of probability'' (132). But there are two forms of Bayesianism - objective and subjective. According to objective Bayesianism, probabilities are real things in nature, while according to subjective Bayesianism, probabilities are the personal degrees of beliefs of agents. Only subjective Bayesianism is consistent with QBism. Moreover, there are other interpretations of probability, such as the \textit{best system's theory}, which resemble frequentism but are more defensible. 

Chapter three is dedicating to explaining QBism and how it resolves the various quantum paradoxes and chapter four specifies various implications of QBism. There is a lot to like about these final two chapters. However, I will focus on the three above mentioned problems: (i) the argument that QBism solves quantum paradoxes is not convincing; (ii) it proves difficult to pin down exactly what QBism says and (iii); as a scientific theory QBism seems explanatorily inert. Let's take each issue in turn. 

\section{Can QBism resolve quantum paradoxes?}

Section 11 ``QBism made explicit'' begins by presenting QBism as a confluence of quantum theory and subjective Bayesianism. QBism is defined by a specific thesis: 

\begin{quote}
``The principal thesis of QBism is simply this: quantum probabilities are numerical measures of personal degrees of belief'' (131).
\end{quote} 

By itself this doesn't suggest a distinctive interpretation of quantum mechanics. For a number of interpretations may help themselves to this principle (and to Bayesianism more generally). For example, in Bohmian mechanics quantum probabilities only arise due to our ignorance of the initial positions of particles. The probabilities may then be interpreted as the degree to which we believe a particle is in a given location. Moreover, some versions of the many worlds interpretation treat quantum probabilities as numerical measures of personal degrees of belief (where the beliefs concern which branch of the multiverse one is in). So we need to know more in order to individuate QBism as a distinctive interpretation. I will return to this problem in the next section. Let's first look to the quantum paradoxes.\\

\noindent
\textbf{The problem of collapse}

The problem of wavefunction collapse is defined as follows: ``there is no mathematical description of how it [collapse] happens in space and time'' (132-3). However, in light of the kinds of dynamical collapse theories presented by Pearle, Penrose, Ghirardi, and others, it seems that mathematically describing collapse is not the real problem. The real problem concerns whether collapse \textit{happens}. You might have thought that this problem will eventually get resolved by the ongoing experimental tests for collapse. But QBists have a different kind of solution:

\begin{quote}
``Qbism solves the problem with ease and elegance. In any experiment the calculated wavefunction furnishes the prior probabilities for empirical observations that may be made later. Once an observation has been made [...] new information becomes available to the agent performing the experiment. With this information the agent updates her probability and her wavefunction - instantaneously and without magic. The collapse sheds its mystery. Bayesian updating describes it and finally makes the missing step explicit'' (133) 
\end{quote}

\noindent
It is unclear how this represents a solution. The assumption seems to be that the concept of collapse arises entirely out of the wavefunction so that just by reinterpreting it, we avoid the need for collapse. But the collapse concept arises first and foremost out of our experimental observations. Consider the double-slit experiment with electrons. If one monitors the slits, one eventually sees two bands of hits on the fluorescent screen. If one does not monitor the slits, one instead sees an interference pattern. Why does monitoring the slits make such a striking difference? Does monitoring the slits cause the states of the electrons to collapse from one kind of state into another? If not, what yields the appearance of collapse? Considerations of how the experimenter updates his probability assignments after seeing such results do not seem relevant to the problem.\\ 

\noindent
\textbf{The problem of nonlocality}

An example is offered to help the reader understand the QBist account of the apparent collapse of entangled particles:

\begin{quote}
``Alice, in New York, picks two playing cards, one black and one red, and tucks them into separate unmarked envelopes, which she seals and then shuffles. [...] She keeps one in her purse and hands the other one to Bob. Alice then leaves the room and travels to Australia. Before she opens her envelope, her degree of belief that Bob has the red card is 50 percent. But upon arrival, as soon as she looks at her own card she knows what's in Bob's envelope twelve thousand miles away, so she updates her degree of belief to either 100 percent or to 0 percent instantaneously. In the meantime Bob's guess about the color of Alice's card, whatever it may be, remains unaffected by her actions. There is no miracle.'' (133-4)
\end{quote}

However, in this classical case we have reason to think that there is a playing card in each envelope at all times during Alice's flight. This partly explains why Alice is right to update her degree of belief about Bob's card instantaneously. So there is no problem with this case. But in the quantum case, we have reason to think that Bob's system did not have any definite value for the measured observable prior to Alice's measurement. So how can the value of Bob's system suddenly become predictable for Alice when she learns the value of her system? This stands in need of explanation. Did Alice's action collapse Bob's system from a non-definite value to a definite value or not? And if it did not, what explains why it appears that it did? As before, this problem is independent of the wavefunction in that we can define the problem in terms of experimental observations alone. Consequently, a subjective Bayesian interpretation of the wavefunction does not seem to be the right apparatus to solve the problem.\\

\noindent
\textbf{The problem of Wigner's friend}

The ``paradox'' of Wigner's friend involves (i) Wigner's friend measuring the spin of an electron and updating his wavefunction accordingly and (ii) Wigner, who turns his back to the experiment and so describes his friend as being in an entangled superposition. Wigner and his friend therefore have distinct wavefunctions. The paradox is then stated as follows:

\begin{quote}
``So who's right? Has the qubit collapsed, or is it still a superposition? As long as the wavefunction is regarded as a real thing or as a description of a real process, the question is no more easily resolved than Bishop Berkeley's infamous question about the tree in the forest: When a tree falls in the forest and nobody hears it, does it make a sound?'' (136)
\end{quote}

But this statement seems incorrect. Theories which admit the reality of the wavefunction yield straightforward answers to this question. For example, it is a trivial implication of collapse theories that Wigner is wrong, his friend is right. Still, the QBist ``solution'' helps us to further pin down what is distinctive about QBism:

\begin{quote}
``According to QBism, there is no unique wavefunction. Wavefunctions are not tethered to electrons [...] they are assigned by an agent and depend on the total information available to the agent. They are malleable and subjective. In short, wavefunctions and quantum probabilities are Bayesian'' (137).
\end{quote}

Here we see a statement that makes QBism distinctive: \textit{wavefunctions are neither real nor represent real physical systems}. This is a striking claim that does not follow merely from quantum probabilities being Bayesian. One is left wondering what the motivation for it is. Wavefunctions are our models of particles that are used to predict the behaviour of those particles. If those models do not (even approximately) represent real physical systems then what does?\\

\noindent
\textbf{The problem of Schrödinger's cat}

How should we represent reality if not via wavefunctions? This problem for QBism is exacerbated when considering macroscopic superpositions, which is the topic of section 12: ``QBism saves Schrödinger's cat''. For if Schrödinger's cat isn't literally in a superposition of being both alive and dead before the box is opened, as the uncollapsed wavefunction prescribes, then what is its actual state? Despite the claim that ``QBism deals with the story as effortlessly as it disposes of the miracle of wavefunction collapse and the paradox of Wigner's friend'' (141), QBism does not appear to have an answer:

\begin{quote}
``QBism refuses to describe the cat's condition before the box is opened and rescues it from being described as hovering in a limbo of living death'' (142).
\end{quote}

Refusing to describe the cat from being in a given state does not seem to save the cat from being in that state. What is the cat's condition? Without an answer we have no guarantee that it is not hovering in limbo.

\section{What does QBism tell us about reality?}

Perhaps the demand to know the physical state of Schrödinger's cat (were we to experimentally realize such a scenario) is asking for too much. Sometimes, when a theoretical problem gets \textit{really} hard, one does not try to solve it, rather, one tries to show that epistemic limitations prevent humans from solving it.  We might call this ``dissolving'' but not solving the problem. Some have tried to dissolve the notorious mind-body problem, for example. Perhaps QBists are only trying to dissolve the quantum paradoxes? 

But we are given no argument for the claim that describing the cat's condition before the box is opened (for example) is beyond us. Section 12: ``The Roots of QBism'' offers quotes from Democritus to Heisenberg, supporting scepticism of ``unvarnished objectivity'' (148). But quotes are not arguments. And the major realist interpretations of quantum mechanics, from collapse theories to many worlds theories to hidden variables theories, to retrocausal theories, all evidence our evolving ability to describe cats in any physically possible situation.

Section 14, ``Quantum Weirdness in the Laboratory'' asserts that ``QBism, by foregoing realism of the Einstein, Podolsky, and Rosen kind, provides a simple, convincing way to avoid spooky action at a distance'' (169). But as discussed, the fundamental problem is found in the lab, where we observe apparent action at a distance. To solve the problem we need a description of what's going on that accounts for the appearances without postulating action at a distance. If QBism is to help us with this problem is must start by telling us what exists. Once we know what exists, we can ask whether those things are local.

Section 15, ``All Physics is Local'', is puzzling. It begins well, by describing how general relativity helped remove the action at a distance inherent in Newton's theory of gravity. It also gives an elegant account of how Feynman diagrams depict the locality of interactions in quantum mechanics. Locality - where interactions only take place at a spacetime point - is a desirable property of physical theories. So we can ask, is QBism a local physical theory? Or to put the question another way:

\begin{quote}
``Where, according to QBism, are the spots, the loci in Latin, at which interactions take place? The black dots of Feynman diagrams are, after all, not actual points in space-time but mere mathematical devices used for calculating probabilities. In plain words, where, according to QBism, does stuff happen?'' (175)
\end{quote}

To answer this question von Baeyer first quotes original QBists, Fuchs, Schack, and Mermin:

\begin{quote}
``QBist quantum mechanics is local because its entire purpose is to enable any single agent to organize her own degrees of belief about the contents of her own personal experience'' (175).
\end{quote}

To clarify, von Baeyer adds:

\begin{quote}
``Personal experiences are recorded (located) in the agent's mind. They follow each other in time but by definition never occur simultaneously in widely separated locations. They are local.'' (175-6).
\end{quote}

But a natural reading of this statement seems in conflict with neuroscience. The cerebral cortex only becomes animated with consciousness when its nerve cells behave in an orchestrated or integrated manner. A conscious experience is hardly a point-like event. But all this talk of personal experience is beside the point. The question of locality is a question about the world not our conscious minds. Is the world local according to QBism or not? Without a description of reality, it seems we must remain agnostic. 

Section 16, ``Belief and Certainty'' attempts to deal with situations in which probability equals unity. The EPR criterion of reality entails that if an electron's wavefunction is a z-spin eigenstate with eigenvalue +1, then the electron \textit{really is} spin-up about the z axis, whether or not it has been measured yet. But then the wavefunction really does describe the electron, at least in this case. QBists respond by reinterpreting probability 1: ``What does it mean when an agent assigns probability 1 to an event? In the context of Bayesian probability, all it implies is that she is very, very sure that it will occur [...] It does not imply anything [...] about the actual makeup of the real world'' (181). I leave it to the reader to ponder this response. 

The final chapter of the book, ``The QBist Worldview'', promises to fill in some details about the QBist ontology. Section 17, ``Physics and Human Experience'', emphasizes how QBism places agents, or users of quantum theory, ``at the center of the action'' (194). This is supposed to be in contrast to traditional physics, which ignores agents completely, and instead just describes atoms in the void. But is QBism guilty of the transpose? 

Section 18, ``Nature's Laws'', compares two conceptions of laws of nature. The first states that laws are real things that govern or guide events in the natural world. The second states that natural events are not governed by laws, laws are just our best summaries of regularities and patterns in events. Qbism adopts the latter. But the summarised events refer to conscious experiences. 

Section 19, ``The Rock Kicks Back'', gives crucial yet cryptic hints about the intended ontology:

\begin{quote}
``According to QBism, a measurement does not reveal a preexisting value. Instead, that value is created in the interaction between the quantum system and the agent [...] the same kind of fact creation occurs when any two quantum systems happen to come together.'' (205-7).
\end{quote}

Everyone should agree that new facts are generated by things coming together, at least in this pedestrian sense: if two particles collide, then the sentence ``two particles are rebounding away from each other'' may come true. 

But I think most would disagree with the first statement and insist that measurements often do reveal preexisting values. For example, if I walk into a room and observe a chair, then the chair was there before I walked in. One wonders whether chairs count as quantum systems in the QBist framework.

QBists appear to want to agree that agents, measuring devices, and measured systems like electrons all exist. We cannot describe the electron, but when an agent interacts with it (via a device), a new fact comes into being, e.g. the fact that the electron is spin-up. 

Fact-creation appears to be the QBist's replacement for collapse. But it looks like this replacement suffers the exact problem that was raised for collapse: ``there is no mathematical description of how it happens in space and time'' (132-3). So it becomes difficult to see what advantage QBism has over the orthodox collapse picture espoused by John von Neumann.

\section{Does QBism explain why the sun shines?}

Section 21, ``A Perfect Map?'', asserts that ``Qbism [...] implies that science is not about ultimate reality but about what we can reasonably expect'' (221), and then argues that this should not be taken as an admission of defeat. Two arguments are offered. Here is one:

\begin{quote}
``Qbism, by moving physics closer to human thoughts and feelings, may have a better chance than raw materialism to solve the ancient enigma of consciousness, the problem of the relationship between the mind and the brain.'' (221).
\end{quote}

But a theory which refuses to describe the physical world is hardly going to be able to describe how that physical world generates consciousness. Here is the other:

\begin{quote}
``QBism doesn't detract in any way from the immense success of quantum mechanics in helping us understand not only the material world but, through biochemistry and neuroscience, the foundations of the life sciences as well. Knowing what we can reasonably expect and how firmly we should expect it is as close as we can come to understanding and controlling the world.'' (221).
\end{quote}

This statement addresses the status of scientific explanantion in QBism. But QBism seems explanatorily inert. For scientific explanations typically explain phenomena in terms of underlying mechanisms. Here is a simple example. Why is the Sun able to produce so much energy over such a long period of time? Physicists tend to answer this question with an explanation that goes something like this: 

\begin{quote}
The sun is composed of many little parts, including hydrogen atoms. If hydrogen atoms fuse together they yield helium. The difference in mass between the products and reactants is manifested as the release of large amounts of energy. According to quantum theory, hydrogen atoms are able to get close enough together to fuse \textit{because they undergo quantum tunneling}. 
\end{quote}

It is not open to the QBist to describe atoms as tunneling, since for QBists tunneling is a psychological phenomenon regarding our degrees of belief. Then what of the explanation? Could the QBist reconstruct the explanation in their own terms? I'm not sure if QBists believe in hydrogen atoms or fusion (as opposed to just the experiences induced by our measurements of them). But assuming they do (the explanations will look worse if they don't!), the worry is that QBist explanations will inevitably look something like this:

\begin{quote}
The sun is composed of many little parts, including hydrogen atoms. If hydrogen atoms fuse together they yield helium. The difference in mass between the products and reactants is manifested as the release of large amounts of energy. According to QBism, hydrogen atoms are able to get close enough together to fuse, \textit{because we expect them to - indeed we are willing to place bets on it}. 
\end{quote}

The worry with the latter ``explanation'' is that it is tautologous: something should be expected to happen because we should expect it to happen. But what we want to know is \textit{why} we should expect it to happen?

\section{Final thoughts}

I have specified some difficulties with QBism. But to end on a positive note, QBism should be applauded as a breeding ground of ideas for multiple disciplines including physics, philosophy, and mathematics, and von Baeyer's book offers an account accessible to all.

The final section, section 22 ``The Road Ahead'', rightly notes that ``One of the most important attributes of a new scientific idea is that it should be \textit{heuristic}, leading to further research, inspiring fresh ideas and questions'' (225). The further research arising from QBism is the ``program of expressing the quantum rules in terms of probabilities rather than wavefunctions''. Von Baeyer discusses how this program has experienced some success and even shows promise in helping to solve problems in pure mathematics. He even speculates that this program ``may turn out to be the basis of a radically new formulation of quantum mechanics without wavefunctions'' (230).

This section was a refreshing one to finish on. It seems that the right way to motivate QBism, is to show that working with it as an assumption can yield unexpected research programs that integrate with other disciplines (e.g. pure mathematics). 

In conclusion, due to the three problems mentioned, this book does not provide a convincing case that QBism is ``The Future of Quantum Physics''. But it does provide an outstanding introduction to two of the key components of QBism (quantum theory and subjective Bayesianism), and places the reader into the mind of the QBist in a way that will aid the ongoing debate over its merit. It is a worthwhile read.
\end{document}